\documentclass[journal]{IEEEtran}
\usepackage{cite}
\usepackage{graphicx}
\graphicspath{ {images/} }
\DeclareGraphicsExtensions{.pdf,.jpeg,.png,.jpg}
\usepackage{float}
\usepackage{tikz}
\usepackage{makecell}
\usepackage{amsmath}
\usepackage{algorithm}
\usepackage[noend]{algpseudocode}
\usepackage{algorithmicx}
\usepackage{array}
\usepackage{url}
\usepackage{booktabs}
\usepackage[flushleft]{threeparttable}
\usepackage{enumitem}
\usepackage{color}
\usepackage{psfrag}
\usepackage{dcolumn}
\usepackage{bm}
\usepackage{hyperref}
\usepackage{xcolor}
\usepackage{comment}
\usepackage{arydshln}
\usepackage{amsfonts}       
\usepackage{nicefrac}       
\usepackage{microtype}      
\usepackage{lipsum}
\usepackage{amssymb}

\usepackage{duckuments} 

\ifCLASSOPTIONcompsoc
 \usepackage[caption=false,font=normalsize,labelfont=sf,textfont=sf]{subfig}
\else
 \usepackage[caption=false,font=footnotesize]{subfig}
\fi

\usepackage{stfloats}

\ifCLASSOPTIONcaptionsoff
 \usepackage[nomarkers]{endfloat}
\let\MYoriglatexcaption\caption
\renewcommand{\caption}[2][\relax]{\MYoriglatexcaption[#2]{#2}}
\fi
\let\MYorigsubfloat\subfloat
\renewcommand{\subfloat}[2][\relax]{\MYorigsubfloat[]{#2}}

\begin{document}

\title{AI-Augmented Behavior Analysis \\for Children with Developmental Disabilities: \\Building Towards Precision Treatment}

\author{Shadi~Ghafghazi, Amarie Carnett, and Leslie Neely,\\ \textit{Child and Adolescent Policy and Research Institute (CAPRI)}\\
Arun~Das and Paul~Rad, \textit{Secure AI Laboratory for Autonomy (AILA)}

\thanks{\copyright 2021 IEEE.  Personal use of this material is permitted.  Permission from IEEE must be obtained for all other uses, in any current or future media, including reprinting/republishing this material for advertising or promotional purposes, creating new collective works, for resale or redistribution to servers or lists, or reuse of any copyrighted component of this work in other works.}

}
\markboth{IEEE SYSTEMS, MAN, \& CYBERNETICS MAGAZINE AUTHOR PREPRINT}%
{Ghafghazi \MakeLowercase{\textit{et al.}}: }
\maketitle

\begin{abstract}
Autism spectrum disorder is a developmental disorder characterized by significant social, communication, and behavioral challenges. Individuals diagnosed with autism, intellectual, and developmental disabilities (AUIDD) typically require long-term care and targeted treatment and teaching. Effective treatment of AUIDD relies on efficient and careful behavioral observations done by trained applied behavioral analysts (ABAs). However, this process overburdens ABAs by requiring the clinicians to collect and analyze data, identify the problem behaviors, conduct pattern analysis to categorize and predict categorical outcomes, hypothesize responsiveness to treatments, and detect the effects of treatment plans. Successful integration of digital technologies into clinical decision-making pipelines and the advancements in automated decision-making using Artificial Intelligence (AI) algorithms highlights the importance of augmenting teaching and treatments using novel algorithms and high-fidelity sensors. In this article, we present an AI-Augmented Learning and Applied Behavior Analytics (AI-ABA) platform to provide personalized treatment and learning plans to AUIDD individuals. By defining systematic experiments along with automated data collection and analysis, AI-ABA can promote self-regulative behavior using reinforcement-based augmented or virtual reality and other mobile platforms. Thus, AI-ABA could assist clinicians to focus on making precise data-driven decisions and increase the quality of individualized interventions for individuals with AUIDD.
\end{abstract}

\IEEEpeerreviewmaketitle

Autism spectrum disorder (``autism") is a developmental disorder characterized by significant social, communication, and behavioral challenges. According to the National Health Statistics Report \cite{zablotsky2020prevalence}, the prevalence of children aged 3-17 years diagnosed with a developmental disability has increased considerably from 6.99\% in 2014-2016 to 17.8\% in 2015-2018. Developmental disabilities can be severe, long-term disorders often including intellectual impairments, physical impairments, or both. Intellectual disabilities, defined by significant limitations in cognition and adaptive functioning, are some of the most common impairments diagnosed during the developmental years, while physical impairments are typically identifiable from birth. Often, these impairments co-occur and individuals diagnosed with autism, intellectual, and developmental disabilities (AUIDD) typically require long-term care and targeted treatment and teaching.

The physical and mental limitations presenting due to AUIDD affect different aspects of life, including personal (self-care, independent living) and social skills (keeping conversations, public speaking), often causing children to learn and develop slower than a typical child. While there is no cure for children with AUIDD, there are several types of treatments such as applied behavioral analysis (ABA), occupational therapy, speech therapy, physical therapy, and pharmacological therapy available. Each of these treatments have proven effective in helping individuals with AUIDD achieve a high level of skill development with earlier treatment leading to larger treatment gains. Detecting and diagnosing these developmental disorders early could help apply the needed treatment procedures and facilitate many students' learning and functioning to improve the behavior of children over time.

\begin{figure}[!t]
\centering
\includegraphics[width=\columnwidth]{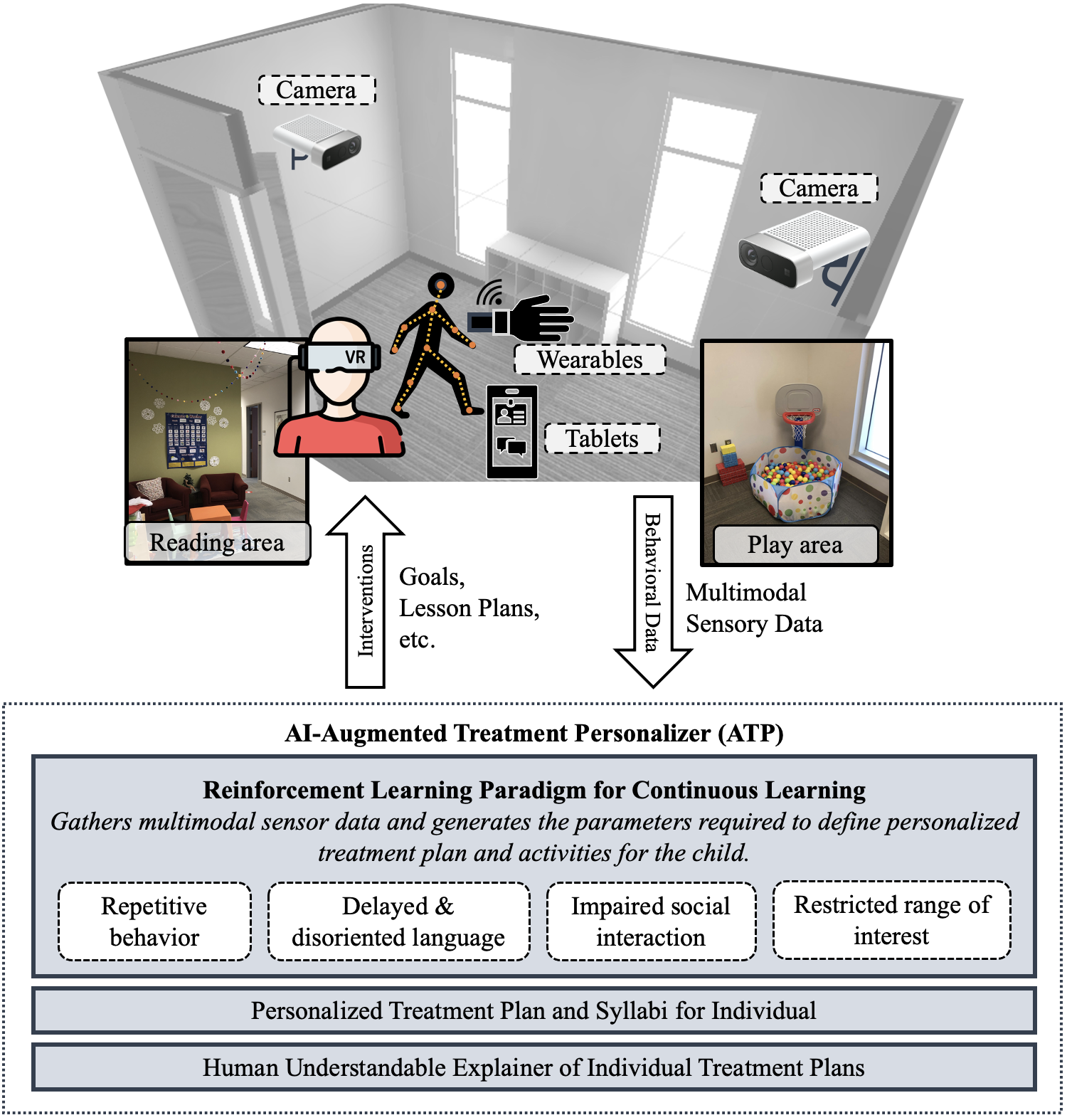}
\caption{System architecture of the AI-augmented applied behavior analytics platforms is illustrated. Multimodal sensory information is collected using both invasive and non-invasive sensors which is processed by AI algorithms to support decision making in treatment and learning paradigms of behavior analysts. All data are stored securely in the cloud accessible by practitioners. Reinforcement paradigms are set up in a personalized fashion unique to each individuals.}
\label{fig:mainarch}
\end{figure}

Effective treatment relies on efficient observation. However, the process of behavioral observation is time consuming requiring clinicians to collect and analyze data \cite{love2012effects}, use the collected data to identify the function of a problem behavior (or what the child is trying to communicate), use collected data to conduct pattern analysis to categorize problems and predict categorical outcomes, use assessment data to hypothesize responsiveness to treatment, design treatment plans using a hypothesis regarding patient responsiveness to treatment, and collect ongoing treatment data to detect the effects of the treatment plan. For example, behavioral assessments used in ABA, such as the functional behavioral assessment (FBA), are used to detect behavior patterns through indirect observations (screening instruments), direct observations and experimental analysis to identify behavioral function. The identified function is then matched with a function-based behavioral intervention \cite{gresham2003establishing}. However, since clinicians rely on human collected data to make decisions, chances of unreliable decisions are likely high, particularly considering varied clinical training that may result in differences from one behavior analyst clinician to another.

Recent research illustrates the importance of personalized treatment plans for individuals with AUIDD. However, while the treatment and education plans may be individualized, they are often not \textit{precise or efficient}. In addition, time spent by clinicians collecting and analyzing the data often detracts from providing empathetic treatment. Given the high impact of digital learning platforms, artificial intelligence (AI), and cloud computing, the purpose of this study is twofold. The First part describes how AI in combination with emerging technologies such as AR or VR in the form of a digital learning platform can benefit more autistic children and provide a personalized adoptive learning paradigm.  The Second part  aims to develop conceptual understanding on how using the efficacy of AI in designing and applying an appropriate assessment framework for ABA and clinical interventions can assist clinicians and educators to effectively assess and monitor each child’s behavior and quickly modify interventions to meet his or her specific needs and to account for various differences across environments.

\section{Augmenting Teaching and Precision Treatment}
Digital platforms and technologies are argued by many to have a pivotal role in the dynamics of changing landscape of clinical treatment. The main arguments include improved support for treatment by 1) contextualizing and increasing motivation of students and promoting engagement through interactive environments and reward structures, 2) providing a learning experience which caters to pace of patient's individual learning, and 3) providing continuous and life-long learning through mobile learning platforms \cite{ng2015change}. 

Several studies have shown promising results in improving the learning performance and boosting motivation to learn using graphical contents and interactions \cite{bacca2014augmented}. Teaching strategies and interventions that utilize digital games in mobile devices or tablets have also shown promise for incorporating behaviors management techniques into games. Restrictive and repetitive behaviors and interests (RRIBs) that might occur while playing games could be monitored and treated using embedded automated redirection to other games or levels to prevent interfering behaviors that prevent access to learning opportunities and help promote calmness \cite{alarcon2018autism}. However, considering the range of functioning for individuals with cognitive disabilities, further studies involving precision of treatment options to ensure individualization are needed, as well as replication of these results across large cohorts of participants \cite{schaefer2016clinical}.

\subsection{Role of Augmented and Virtual Reality}
Emerging technologies such as augmented reality (AR), including virtual reality (VR) and mixed reality (MR), is in the forefront of recent technology-embedded practices that overlays reality and supplies additional layers to augment the perception of users \cite{azuma2001recent} as well as enabling real time interaction of real and virtual objects \cite{parsons2017potential}. Recent  interest in using AR and VR technologies to aid adults and children with ASD provides additional sensory information such as eye- tracking as well as a virtual platform to continuously interact with people and environment around them in a controlled setting while collecting data for future analysis. 

Safe and side-effect-free technologies are changing how AR/VR platforms are being found to be beneficial for improving soft skills, behaviors, and improving emotional skills \cite{sahin2018safety}. However, the importance of personalized services to provide augmentation for individual learners has yet to be researched at large. Thus, the need for personalized adaptive learning paradigms is required to improve engagement, autonomy, and to promote individualized preferences for children with cognitive disorders. Data-driven algorithms could make use of these digital technologies to improve data collection while reducing the demands placed on behavioral analysts and educators and the time-required to collect and label these behavioral data. 

\subsection{Artificially Intelligent Methods in Behavioral Health}
Recent advancements in artificial intelligence (AI) have enabled real time human action performance \cite{bendre2020human}, facial behavioral analysis \cite{baltrusaitis2018openface}, speech analysis \cite{memari2020speech}, speech disfluency detection \cite{das2021interpretable}, stereotypical motor movement from sensory data \cite{rad2018deep}, many more. Published research from the last 5 years shows the use of a wide variety of sensory inputs to predict human behavior, diseases, and cognitive states using AI methods, especially deep learning (DL). Electroencephalograms (EEG) has been used extensively to study the internal brain states by recording the electrical activity of brain waves for health monitoring \cite{pantelopoulos2009survey}, predicting diseases such as Parkinson's \cite{oh2018deep}, and assessing emotional disorders \cite{lin2017eeg}. Even though using EEG to study Autism could have contradictions based on the experimental conditions during EEG registration between subjects, age differences, and diversity of subjects, the abnormal EEG laterization in subjects with ASD can be leveraged to build AI models to predict traits of autism \cite{penchina2020deep}.

\begin{figure}[!b]
\centering
\includegraphics[width=\columnwidth]{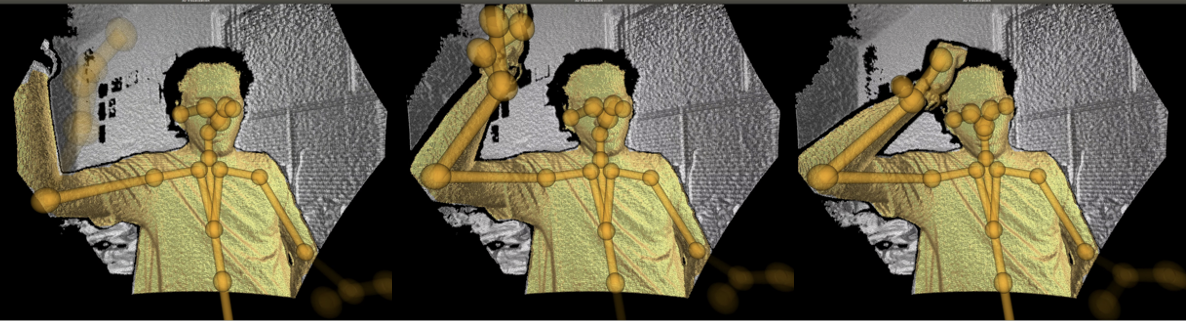}
\caption{Tracking aggressive behavior using body tracking AI algorithm is illustrated. Here, forceful movements to the head can be classified as banging the head using a temporal action indexing module \cite{bendre2020human} as described in our previous research.}
\label{fig:actionintensity}
\end{figure}

Prior research using deep learning algorithms illustrate the successful use of facial videos collected using cameras to estimate the attention and engagement of children with developmental disorders \cite{di2018deep}. Similarly, inertial measurement unit (IMU) sensors which rely on accelerometer, gyroscope, and magnetometers that can collect information about the frequency, intensity, and duration of physical activities have shown to detect stereotypical movements in ASD children \cite{rad2018deep}. As our children rely more on digital devices such as iPads and mobile phones to read, learn, and interact from an early age \cite{linke2017tabooga}, it is only natural to research in facilitating access to individualized digital content through a variety of interfaces (such as games, interactive lessons, maps, and more) to cater for the exceptional learners.

\begin{table*}[ht]
\caption{Summary of common behavioral challenges, their descriptions, and possible measurements using sensors}
\label{tab:behaviorsandsensors}
\centering
\begin{tabular}{||p{2.7cm}|p{7.5cm}|p{6.4cm}||}
\hline
Behavioral Challenge & Description & Possible Sensors and Measurements \\ 
\hline\hline

Communication Delays	
& Children with neurodevelopmental disabilities, such as ASD, often require communication interventions. Additionally, approximately 30\% of individuals with ASD do not fully acquire spoken communication \cite{wodka2013predictors}.
& Vocal communication could be more precisely evaluated using high tech data collection, which could promote more accurate and reliable intervention procedures. Automated data collection could be embedded in speech-generating device software along with isolated teaching platforms to teach the use of these alternative communication modes and promote data-based decision making.\\

\hline
Problem Behavior 
& End result of a chain of events that usually triggers (reactive, impulsive) during interactions with social and physical environments often accompanied by anger or frustration. Problem behavior, such as aggression, can be towards self, others, or the environment, while verbal aggression is usually towards others. Understanding the underlying conditions and patterns are very important for BA’s to help reduce problem behavior.
& Problem behavior such as biting, hitting, kicking could be tracked using body tracking sensors and algorithms \cite{bendre2020human}. Vocal aggression such as screaming, shouting, use of foul language, etc. could be found by recording the speech and studying them using intelligent algorithms.\\

\hline
Agitation 
& Restless or hyperactive stages, usually due to unrecognized physical or sensory discomfort, which could lead to anxiety, verbal or physical aggression.
& Physiological data from wearable devices and EEG systems could help understand the long-term triggers and etiological factors leading to agitation.\\

\hline
Vocal Stereotypy 
& Persistent repetitions of words, phrases, or sounds without contextual or functional meaning in the current social setting.
& The vocal behavior could be found by recording the speech and studying them using intelligent algorithms in a functional analysis experimental setting. \\

\hline
Sleep Disorder 
& Trouble falling asleep or staying asleep for long periods of time. Could affect academic and social behaviors due to mood problems, memory, concentration, and learning problems, sluggish reaction time, etc. caused by disrupted sleep patterns, insomnia, or sleep apnea \cite{carnett2020quantitative}. 
& Wearable sleep staging sensors or sleep sensing mats could identify sleep patterns before and after behavioral therapies to study impact of the therapies in sleep and to design therapies that could improve sleep.\\
\hline

Social Delay 
& Difficulties with understanding social rules and interactions. Can be related to communication delay and engagement in problem behavior but may also be present in individuals without communication delays.
& Specially designed games on AR/VR environments could teach social skills and provide a structured environment for practicing vital social interactions.\\

\hline

\end{tabular}
\end{table*}

\begin{figure*}[!t]
\centering
\includegraphics[width=0.9\textwidth]{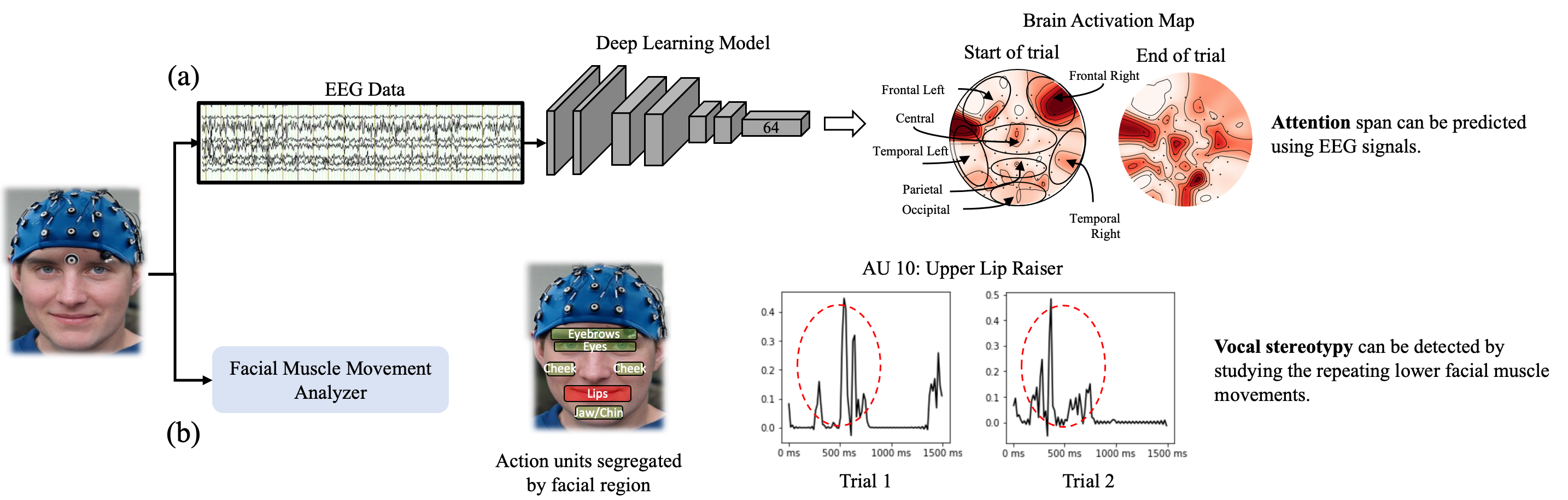}
\caption{Attention span of children can be studied by capturing EEG signals or eye-tracking information whereas vocal stereotypy can be predicted by studying the facial muscle movement patterns over time. In (a), we illustrate a deep learning model generating a prediction and corresponding brain activation maps highlighting the most attributing region of the brain. Vocal stereotypy (b) can be detected by observing the repetitive motion in the lower facial muscle groups using a facial muscle movement analyzer.}
\label{fig:attnstereo}
\end{figure*}

\subsection{Challenges of AI in Behavioral Health}
Despite the impressive role of AI in behavioral health, there are two key open challenges which limits its use in clinical decision-making. They are: 1) limited amount of labelled data to train AI algorithms and 2) black-box nature of deep neural network models. There are two potential solutions to these problems. Firstly, self-supervised representation learning has been recently used to learn meaningful dense representations from small amount of data. Also, reinforcement learning paradigms can learn to optimize based on an exploration-exploitation paradigm on any defined environment. Secondly, explainable artificial intelligence (XAI) methods can be used to improve the transparency and trust of decision-making by generating meta-information to describe `why' and `how' a decision was made while suggesting `what' features influenced the decision the most \cite{das2020opportunities}. 

\section{Personalized Explainable AI to Improve ABA and Treatment}
ABA is an effective and evidence-based treatment to address individual needs for children with developmental and intellectual disabilities.  Additionally, ABA treatment programs have been successfully implemented in different settings such as home, communities, school, and other educational centers. The beauty of ABA is that it focuses on data-based decision making, which allows the clinician to tailor interventions and personalize treatment options to an individual’s unique situation and need. However, as previously discussed, the analysis process is time-consuming and has limitations due to its reliance on humans as the data collectors and pattern analyzers, which is subjective given the reliance on professional judgment. These limitations can also be confounded by the ever-increasing demands placed on behavior analysts and educators, such as increased caseloads, time demands, and limited resources.
Augmenting and complementing the systematic evaluation of subject’s data using intelligent algorithms is an avenue that warrants further research. Recent literature highlights the importance of online treatments and large adoption of telehealth solutions [24] across the globe. 

Recently, artificial intelligence (AI) has been widely used to investigate autism with the overall goal of simplifying and speeding up the diagnostic process as well as making access to early interventions possible \cite{marciano2021artificial}, in which supervised machine-learning algorithms have been studied as useful aids to support decision making in screening \cite{song2019use}, diagnosing autism \cite{sadiq2019deep}, shorten diagnosing time \cite{wall2012use},  predicting behavior and constructing predictive models \cite{linstead2015application,thabtah2019machine}. More recently, there has been a number of developments demonstrating the feasibility of automated facial behavior analysis systems for identifying and pre-diagnosing of autistic individuals. This research has shown positive advancements in designing the AI based platform algorithm to detect and evaluate autistic patients based on brain activation patterns and mental disorders \cite{heinsfeld2018identification}, faster screening \cite{wall2012use}, skill assessment \cite{pandey2020guided}, social interaction analysis \cite{bekele2016multimodal, zhao2018hand}, facial expression \cite{tsangouri2016interactive,haque2019facial} and eye tracing \cite{dalmaijer2014pygaze}. However, few studies have considered the feasibility of AI in a natural setting, such as homes or schools \cite{vijayan2018framework}.  Effective treatment of AUIDD relies on efficient and careful behavioral observation and assessment conducted by board certified behavioral analysts (BCBA) through assessment, both direct and indirect. Although current practices for ABA treatment are highly effective, they are often time consuming and are subjective to human errors.

Building an Explainable AI-Augmented Learning and Applied Behavior Analytics (AI-ABA) platform could complement licensed behavior analysts and therapists who rely on direct observation of audio-visual cues and other physiological data available during a session to diagnose and provide feedback to subjects. AI-ABA could make use of facial expressions, extremity movements, speech tone, heart-rate, and other available data to build automated pipelines to detect, diagnose, and alert BA’s of during treatment sessions with clients. Some of the desirable qualities of the AI-augmented ABA and learning platform are as follows:

\begin{itemize}
  \item Ability to collect cognitive, perceptive, speech, movement, and physiological multimodal data of individuals.
  \item Ability to define systematic experiments with dynamic reinforcers in virtual or physical environments.
  \item Ability to manage experimental data and results of recurring experiments of individuals.
  \item Provide early detection of behavioral changes in children and explain why, how, and what features were used for the prediction.
\end{itemize}

Moreover, AI-ABA platforms can promote self-regulative behavior using reinforcement-based AR/VR or game environments and collected multimodal data. It could also promote creativity and curiosity by designing personalized dynamic environments which involve physical activities and speech interactions while completing set tasks. By supporting games and virtual agents, AI-ABA could promote the development of spoken communication and social communicative behaviour such that the collaboration and social skills can be improved by interacting with virtual agents in social settings using natural language technologies.

Figure \ref{fig:mainarch} illustrates the system architecture of an AI-ABA paradigm which complements human intelligence with AI algorithms. General system architecture of AI-ABA must consist of 1) multimodal sensing technologies to collect a variety of sensor data, 2) AI-augmented Treatment Personalizer (ATP) to ingest the sensor data and generate personalized treatment plans and IEPs to children, 3) suite of explainable AI algorithms to support various requirements of the ATP, and 4) integration to various presentation formats or front-end technologies such as AR/VR/MR \cite{bilyalova2019digital} or tablets \cite{kucirkova2014ipads}. 

By designing hardware-software systems which caters to the need of behavior under study, AI-ABA could be used as a platform for closed-loop real-time multimodal data analysis. For example, data repetitive behaviors in individuals with AUIDD could be collected using cameras or other sensors. These collected data can be streamed to a compute device which can host AI algorithms which are optimized run on limited hardware resources using quantization operations \cite{kwasniewska2019deep}. The AI predictions could be used to initiate a behavioral correction algorithms which are either deterministic or not. By exposing the results of behavioral correcting algorithm to specific web Application Programming Interfaces (API's), these results can be viewed on mobile or desktop devices or visualized on a monitor screen. In the following section, we summarize a few behavioral challenges of individuals with AUIDD and how AI-ABA could be a beneficial technology in augmenting treatment and teaching.

\subsection{Behavioral Challenges and Sensing Technologies}
Table \ref{tab:behaviorsandsensors} summarizes some of the common treated behavioral challenges for children with AUIDD and possible sensor measurements that could improve data collection while reducing the physical and mental burden of arduous observational data collection for BA's during the treatments. Agitation, aggression, and stereotypy seen in children with developmental disorders could reoccur due to the process of reinforcement that parents unknowingly encourage in the household. Hence, it is very important to track behavioral changes in multiple settings and environments to design individualized treatment plans or IEPs for children. 

A problem behavior or challenging behavior is any culturally abnormal behavior that could jeopardize the physical safety   of the individual or others that often restricts the person from social or communal activities. Self-harm or self-injurious behavior (SIB), harming caregivers, or general aggression is seen in children with low frustration tolerance. Throwing items at people, hitting themselves with objects, banging head against an object, etc. are visible behaviors of aggression and agitation in children with problem behaviors. Audible and/or visual behavior such as vocal stereotypy is commonly identified by parents as their children repeat words repetitively. However, despite the many intervention models described in research, only a few focuses on the impact of intervention model in vocal stereotypy and the secondary impacts in other behaviors \cite{ahearn2007assessing}.

\begin{figure}[!t]
\centering
\includegraphics[width=\columnwidth]{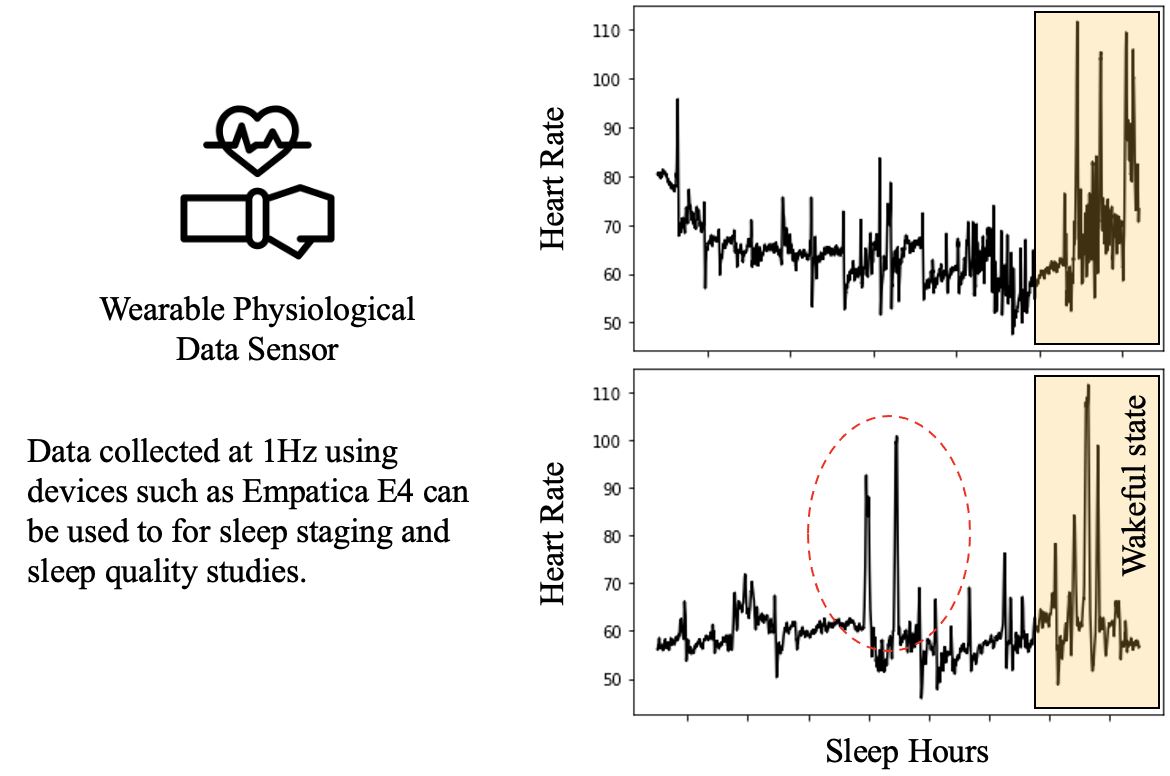}
\caption{Wearable physiological sensors such as Empatica E4 can be used to collect heart rate data and study variability to carry out sleep staging and sleep quality analysis before and after therapy sessions. Here, the red dotted lines indicate intermittent waking up states.}
\label{fig:sleep}
\end{figure}

Problem behaviors can be evaluated in multiple settings and environments by collecting multimodal sensor data including EEG, gait and limb movements and body tracking using cameras. Figure \ref{fig:actionintensity} illustrates one such method to index mild versus intense actions in humans. By generating a temporal map of body movements, a spatio-temporal LSTM based attention network is used to highlight the areas of the body with rapid movements to classify actions. A kinetic fuzzy intensity analysis network generates an action intensity based on the temporal action map and deep learning prediction. Attention and agitation can be studied by a temporal analysis of the facial muscle movements and emotional states. 

Attention span can be studied using a temporal analysis of EEG signals as illustrated in Figure \ref{fig:attnstereo} (a). Brain attention maps can be generated using deep learning algorithms to understand the pre and post brain states leading to a more detailed understanding than traditional methods. Vocal stereotypy can be detected by observing the lower facial muscle movements for repetitive motions. This is illustrated in Figure \ref{fig:attnstereo} (b). Here, the microexpressions on the face leading to repeated behaviors and sounds can be used as inputs to a deep learning algorithm to classify vocal stereotypy among other behaviors. We have found Microsoft Azure Kinect camera to be a well-rounded sensor to record body tracking data, RGB, IR, and depth videos, and also, stereo audio data. Additionally, we found Empatica E4 wearable sensor to be reliable to study sleep disorders using heart rate, electrodermal activity, and other sensory data as illustrated in Figure \ref{fig:sleep}. Integration of data collected via behavioral sensing can lead to deeper understanding of the triggers of problem behavior, such as poor sleep the night before. It can also help to distinguish behaviors triggered by environmental events, versus those rooted in biological processes, and facilitate early identification of behaviors via precursors, such as heart rate. Precise understanding of behavioral triggers is essential to ensuring effective and efficient treatment.

Various other actions, such as pacing in a repeated circular motion, body rocking, back and forth movement of fingers, tapping objects repeatedly, etc. are used as pointers to understand the hidden behaviors that might help with a diagnosis. Single-subject research designs (SSRDs), focused on individual participants, are often used, even now, to identify and evaluate interventions due to the heterogeneity of the AUIDD population. However, integration of AI into ABA can allow for collection of larger data sets and analysis of multi-modal behavioral data, enhancing research standards and the strength of the research evidence. In reality, finding interventions that can be considered evidence-based practices (EBPs) require large cohorts of subjects and multi-session datasets [28]. Hence, scalable architectures such as AI-ABA should be explored as a means to collect data and to promote early detection AUIDD from large subject populations and predictive models facilitating responsiveness to intervention.

\section{The Road Ahead}

Systems such as AI-ABA could assists practitioners in making the  precise data-based decisions to increase the quality of individualized intervention for individuals with AUIDD. This type of system is likely to have profound effects on the improvement in treatment efficacy and treatment outcomes, through a blend of both real-world and virtual elements that embeds generalization of targeted skills across multiple environments for the various aspects of an individual's life. Recent studies illustrate the importance of active involvement of children in the intervention process \cite {singh2015interactions,alias2017analysis}. Additionally, providing an learning environment with effective interaction and communication is another important factor that should be taken into account \cite{csafak2015siblings,syrjamaki2017enhancing}. 
Since AI augmented ABA provide the optimal use of delivering ABA services for children with developmental and cognitive needs, the system could easily be integrated to  across environments, such as schools and home to deliver a more comprehensive treatment program. This smart and connected health via behavioral sensing would not only be highly desirable for existing outpatient clinics, but can be integrated into telehealth platforms to facilitate service access to those living in rural and deprived services areas, and can provide a level of system resiliency during periods where human interaction is limited, such as the ongoing COVID-19 pandemic. In addition, applying home-based ABA can help ensure parents involvement with their child training alongside other stakeholders (e.g., teachers, caregivers) to create a more comprehensive support network for service delivery.

AI-ABA will help researchers to focus on the precision within intervention research, reinforcers effectiveness, and individualized treatment models while augmenting part of the data collection and analysis to AI algorithms. For example, a system capable of handling multiple sensory inputs for data capture could in a plug-and-play manner collect data using specific application programming interfaces (APIs) and process them using existing machine learning algorithms. The processed data and results could be used to dynamically influence the virtual environments, learning structures, the precision of treatment plans, effectiveness,  and the incorporation of AR or VR digital platforms to promote greater access to intervention and generalization of treatment effects. AR and VR feedback loops could increase learning engagement and raise comprehension of topics, provide interaction, improve communication, trigger imagination, and enhance problem-solving skills, especially when involving spatial skills \cite{bilyalova2019digital}. A combination of AR and VR technologies with other invasive and non-invasive data collection systems could collect both physiological and behavioral data to study temporal dynamics of behavior in children. Additionally, this could enable just- in-time adaptive interventions (JITAIs) by collecting precise data and provide a repository of prior behavior of each client.

\section*{Acknowledgment}
This project was funded partly by the Open Cloud Institute (OCI) at University of Texas at San Antonio (UTSA) and partly by the UTSA Brain Health Consortium and Office of the Vice President for Research, Economic Development, and Knowledge Enterprise. Arun Das and Shadi Ghafgazhi contributed equally. The authors gratefully acknowledge the use of the services of Jetstream cloud.

\section*{About the Authors}

\textbf{\textit{Shadi Ghafghazi}} (sh.ghafghazi@gmail.com) is currently a Senior Lecturer at the University of Applied Science and Technology where she is teaching special needs education courses. She earned her Master's degree in Information Science and Knowledge Studies from Shahid Beheshti University and reaped her Bachelor's degree in Library and Information Science from University of Tehran, Iran.

\textbf{\textit{Amarie Carnett}} (amarie.carnett@vuw.ac.nz) is a Senior Lecturer with the College of Education, Victoria University of Wellington, New Zealand and a Research Faculty at the University of Texas at San Antonio. She is a doctorate level Board Certified Behavior Analyst (BCBA-D).

\textbf{\textit{Leslie Neely}} (leslie.neely@utsa.edu) is an Associate Professor in the College of Education and Human Development, University of Texas at San Antonio and Director of the Child and Adolescent Policy and Research Institute. She is a doctorate level Board Certified Behavior Analyst (BCBA-D).

\textbf{\textit{Arun Das}} (arun.das@utsa.edu) is currently a Ph.D. candidate at UT San Antonio, TX, USA where he focuses on explainable AI and self-supervised algorithms for healthcare and neuroscience domains. He earned his MS in Computer Engineering from UT San Antonio, TX, USA and his B.Tech. from Cochin University of Science and Technology, Cochin, India. Arun is a graduate student member of the IEEE Lone Star section and the IEEE Eta Kappa Nu honor society.

\textbf{\textit{Paul Rad}} (paul.rad@utsa.edu) is an Associate Professor with the Department of Computer Science, The University of Texas at San Antonio. He is a Senior Member of the National Academy of Inventors and a Senior member of IEEE.


\bibliographystyle{IEEEtran}
\bibliography{references}

\end{document}